\documentclass{sf2a-conf2024}
\usepackage{graphicx}
\usepackage{hyperref}
\usepackage[]{natbib}  
\usepackage{epstopdf}

\def\BibTeX{{\rm B\kern-.05em{\sc i\kern-.025em b}\kern-.08em
    T\kern-.1667em\lower.7ex\hbox{E}\kern-.125emX}}
\bibpunct{(}{)}{;}{a}{}{,}  %%%%%%%%%%%%%  A&A bibliography style
\begin{document}

\TitreGlobal{SF2A 2024}

%%-----------------------------------------------------------------
%%      the top matter
%%

\title{What trade-off for astronomy between greenhouse gas emissions and the societal benefits? A sociological approach}

\runningtitle{What trade-off for astrophysics?}

\author{P. Hennebelle}\address{Universit\'e Paris-Saclay, Université Paris Cité, CEA, CNRS, AIM, 91191, Gif-sur-Yvette, France }

\author{M. Barsuglia}\address{Universit\'e Paris Cit\'e, CNRS, Astroparticule et Cosmologie, F-75013 Paris, France}

\author{F. Billebaud}\address{Laboratoire d’Astrophysique de Bordeaux, Univ. Bordeaux, CNRS, B18N, allée Geoffroy Saint-Hilaire, 33615 Pessac, France}

\author{M. Bouffard}\address{Nantes Universit\'e, Université d’Angers, Le Mans Universit\'e, CNRS, Laboratoire de Plan\'etologie et G\'eosciences, LPG UMR 6112, 44000 Nantes, France}

\author{N. Champollion}\address{Universit\'e Grenoble Alpes, CNRS, IRD, INRAE, Grenoble-INP, Institut des G\'eosciences de l’Environnement (IGE, UMR 5001), 38000 Grenoble, France
}

\author{M. Grybos}\address{E2Lim (UR 24 133), Universit\'e de Limoges, 123 Avenue Albert Thomas, CEDEX, 87060 Limoges, France}

%\author{A. Hardy}\address{Centre Emile Durkheim, CNRS, Science Po Bordeaux, CNRS, Bordeaux, France}

\author{H. Meheut}\address{Universit\'e C\^ote d’Azur, Observatoire de la Côte d’Azur, CNRS, Laboratoire Lagrange, Nice, France}

\author{M. Parmentier}\address{Institut de Génomique Fonctionnelle (IGF), Universit\'e de Montpellier, Centre National de la Recherche Scientifique (CNRS), Institut National de la Sant\'e et de la Recherche Médicale (INSERM), 34094 Montpellier, France}

\author{P. Petitjean}\address{Institut d’Astrophysique de Paris, Sorbonne Universit\'e and CNRS, 98bis boulevard Arago, 75014 Paris, France}

%\author{J.-P. Author2}\address{Institute XYZ, 1299 City, OtherLand}

%% IF Author3 has the same affiliation than Author1:
%\author{C.\,E. Author3$^1$}

%% IF Author3 has its own affiliation:
%\author{C.\,E. Author3}\address{Dept. of Chess, University of Games, 35101 Las Vegas, Monaco} 

%% IF Author3 has two affiliations, the one of Author1 and a second one:
%\author{C.\,E. Author3$^{1,}$}\address{Dept. of Chess, University of Games, 35101 Las Vegas, Monaco} 

%% Keep this line, even if the page will be settled afterwards.
\setcounter{page}{237}

%%-----------------------------------------------------------------

\maketitle

%%-----------------------------------------------------------------
%%        The abstract
%% 
%%  Warning!  within the abstract:
%%  - do not use macros. 
%%  - do not use commands like: \cite, \citet, \citep ... etc.

%\begin{abstract}
%The threat posed  to humanity by global warming  has led  scientists to question the nature of their activities and the need
%to reduce the greenhouse gas emissions from research. Until now  most studies aimed at quantifying the carbon footprints and relatively 
%less works have attempted to discuss the utility of research activities and the necessary trade-off that scientists must make between 
%social utility and GHG emissions. Here we present the results of 28 semi-directive interviews of french astrophysicists, with the aim of 
%understanding how astrophysicists perceive these questions. Our most important findings are that in most cases, astronomy is considered 
%to have a social impact mainly regarding education but also for the fascination astronomical questions exert on the general public. The reduction of GHG emissions are 
%believed to be necessary and most often reductions at a personal levels have been achieved. The question of community-wide reductions , and 
%in particular the possible reductions of large facilities is much more mixed. Semi-directive interviews appear as a powerful tool, complementary to quantitative surveys, to understand the choices that need to be made by scientific communities in the context of global warming.
%\end{abstract}

\begin{abstract}
The threat posed to humanity by  global warming has led scientists to question the nature of   their activities and the need to reduce the greenhouse gas emissions from research. Until now, most studies have aimed at quantifying the carbon footprints and relatively less works have addressed the ways GHG emissions can be significantly reduced. A factor two reduction by 2030 implies to think beyond increases in the efficacy of current processes, which will have a limited effect, and beyond wishful thinking about large new sources of energy. Hence, choices among research questions or allocated means within a given field will be needed. They can be made in light of the perceived societal utility of research activities. Here, we addressed the question of how scientists perceive the impact of GHG reduction on their discipline and a possible trade-off between the societal utility of their discipline and an acceptable level of GHG emissions. We conducted 28 semi-directive interviews of French astrophysicists from different laboratories.  Our most important findings are that, for most researchers,  astronomy is considered to have a positive societal impact mainly regarding education but also because of the fascination it exerts on at least a fraction of the general public. Technological applications are also mentioned but with relatively less emphasis. The reduction of GHG emissions is believed to be necessary and most often reductions within the private-sphere have been achieved. However, the question of community-wide reductions in astrophysics research, and in particular the possible reductions of large facilities reveals much more contrasted opinions. In conclusion, semi-directive interviews appear as a powerful tool, complementary to quantitative surveys, to understand  the attitudes of scientists with respect to the trajectories and  the choices that need to be made by scientific communities in the context of global  warming.
\end{abstract}

%% Insert the keywords (to appear in the ADS indexing)
%% Keywords must be separated by a comma
\begin{keywords}
global warming, survey, semi-structured interview, societal benefits, greenhouse gas emission, trade-off
\end{keywords}

%%-----------------------------------------------------------------

\section{Introduction}
%%---------------------
According to the 6th IPCC assessment report (IPCC 2021), it is now firmly established  that
human influence, and more precisely greenhouse gas emissions (GHG), are triggering unprecedentedly fast  global warming. 
To limit the increase of the temperature below 1.5-2$^o$ C, it is necessary to drastically reduce the GHG emission as for instance 
defined by the French {\it Strat\'egie Nationale Bas Carbone} 
\footnote{\url{https://www.ecologie.gouv.fr/politiques-publiques/strategie-nationale-bas-carbone-snbc}}
and by the European Green Deal
\footnote{\url{https://commission.europa.eu/strategy-and-policy/priorities-2019-2024/european-green-deal_en}}.
The required effort, which involves halving GHG emissions in 2030, is very challenging and will only succeed if all sectors of  society take an active
role in the process. In the Astronomical research community, this has triggered several studies \citep[e.g.][]{2020NatAs...4..843S} which have attempted to quantify the carbon footprint of large astronomical meetings
 \citep{2020NatAs...4..823B,2023NatAs...7..244W}, astronomical institutes
\citep[e.g.][]{2020NatAs...4..812J,2021NatAs...5.1195V,2022NatAs...6.1219M} or the construction and maintenance of large facilities
\citep{2022NatAs...6..503K}. The latter in particular has been found to be the major contributor of GHG emissions in astronomy. 
%\citep{2023NatAs...7..244W} 
 Generally, it has been found that purchases dominate the carbon footprint of laboratories  \citep{DePaepe2023.04.04.535626} and this 
 has important implications regarding the attitude scientists may adopt when comes the decision to reduce the carbon footprint of their activities. 
 Therefore, a significant reduction of the GHG emissions of the
  research activities, i.e. on the order of a factor of  about two, would inevitably require significant cuts in purchases, particularly in large facilities like telescopes and satellites in astrophysics.  This would almost certainly impact our present way of doing astronomy.
  Moreover, like all human activities, academic research and in particular astrophysical research is bringing various benefits to human societies, 
   this suggests that some kind of trade-off must be established between GHG emissions and the social benefits of astrophysical research. The latter however are not easy to assess.
Indeed  the importance of the roles, scientists can play in our societies remains subjective. 
Whereas numerous studies have quantified the carbon footprint of research activities, far fewer, have addressed
 the societal and human aspects of the climat crisis within academia  \citep[e.g.][]{blanchard:halshs-03567136} in spite of their importance to achieve actual emission reductions
\citep{ ragueneau:hal-04679110}.

The present contribution  attempts to address these issues using the methods of sociology. We present the results of a series of interviews
of astrophysicists, who have been asked to answer   questions related to how they perceive the utility of astrophysics and what would be 
the impact on astrophysics if serious reductions  would occur.  The present paper is a preliminary overview of a scientific work currently being finalized (Hardy et al., 2024, in prep).
In the second section we describe our methods. The third section presents  results regarding the societal benefits of astronomy. 
Fourth section is devoted to the reduction themselves and their perceived consequences. Fifth part concludes the paper.

\section{Methods}
%%-------------------------
This study was conducted by a group composed of astrophysicists, biologists, geoscientists and sociologists as a part of the collective Labos1point5  \footnote{\url{https://labos1point5.org/}}. We carried out 28 semi-structured interviews with french  research staff coming from about 10 astrophysical institutes, between June 2022 and June 2023.
Such qualitative methods are complementaries from quantitative surveys \citep[e.g.][]{blanchard:halshs-03567136}.
 Whereas most research-participants were colleagues to us, we were generally unaware of their knowledge and opinions regarding climat issues, carbon footprint and 
 the reduction of emissions. Since we had only few refusals, it seems unlikely that our sample was significantly biased in a sense or another. 
  In a first phase, 14 interviews have been conducted. Our grid of interviews 
consisted in 9 questions  
that were designed firstly to determine participants' level of knowledge of the climate crisis, and their commitment to the environmental crisis in general. Secondly, they sought to determine the perception of the societal usefulness of astrophysics. Finally, the consequences of a significant reduction of the emissions induced by participants' personal research, by  their institutes and by the discipline as a whole were posed. 
The exchanges lasted between half an hour to one hour. 
In a second phase, another set of 14 interviews have been carried out. The grid was much more detailed and the interviews were lasting, on average, about one hour.  This second phase 
allowed  to test our first results and  to verify whether “data saturation”\citep{doi:10.1177/1525822X05279903} was reached. All interviews have been carried out online, after which they have been transcribed using the software Transkriptor. They have then been coded and analysed. The research-participants are evenly distributed between man and woman and they are aged between  25 to 68 years. They represent different fields of  astrophysics and have various professional statuses (junior and senior researchers from CNRS or universities, technical engineers), including three non-permanent and three retired scientists.  A possible bias of our study is that it has been conducted exclusively in France and that there maybe some cultural specificities. Similar studies need 
to be undertaken in other countries.
The verbatims were translated from French into English using deepl software.

\section{Societal benefits of astronomy}
When asked about the usefulness of astronomy toward the society, most interviewees responded positively and several categories were mentioned. Four of them 
which have appeared repeatedly during the interviews are detailed below.

\subsection{Research drives technology}
One   category of utility  often mentioned and generally considered obvious in the interviews, are the technological applications which are consequences of various  developments
 made in astronomy. It is worth stressing that it has been always regarded as a positive consequence of astronomical research, that is to say the possible negative impacts 
 that these technological developments may have on the environments are not mentioned. \\

{\it "So yeah, there's the usual thing, I'd say that even though it's fundamental research, there are always some technological applications. Even if they're not direct, there are technological applications, when we develop imaging, when we develop uh... we miniaturize something that's going to be sent up in a satellite, I know that there have been uh... concrete research applications, that's all. But that's not the fundamental goal. And after that, the fundamental aim is to gain a better understanding of our Universe." } \\

This is not necessarily restricted to real experiments but also holds for theoretical works and numerical simulations for instance. \\

{\it "We have experiments that involve cutting-edge technology, so the link with industry is clear. Beyond instrumentation, we also have everything that relates to theory, and with numerical calculations, here there's also a clear link with industry. So the link with industry is obvious."}

\subsection{Spreading the message that colonizing other planets is unrealistic}
Another important message, that is sometimes mentioned, is that colonizing other planets is not a realistic solution to the climat crisis. 
Obviously we now know that whereas there is a handful of planets in the Universe, reaching them is, and will remain for a long time, simply impossible. \\

{\it  "We can get the message across that, yes, there are lots of exoplanets, but there's only one Earth, so we'll never be able to get to the others, so I think we have to stay visible..."  } \\

It is suggested that the "no planet B" warning may be used as an invitation to care more about Earth preservation, not only for us but also for life in general.  \\

{\it "Perhaps the only thing that... that I'm involved in that I think is important from an astronomical point of view, is to spread the message that there is no Planet B. Because one of the messages is to say, well, be careful. Because one of the messages is to say, well, we've got to preserve the planet, because that's going to make human life, and even life in general, difficult on the planet. The planet isn't going to disappear, but any form of life is likely to be a bit problematic if we're not careful." }

\subsection{Astronomy as a powerful vehicle for science education}
A very important area where astronomy is almost always mentioned as playing a significant, if not drastic role, is science education. The fascination that 
astronomy is exerting on students for instance is mentioned as a key to attract them. 
Whereas astronomy may contribute to specific scientific fields like climate studies, its border impact on maintaining curiosity and imagination, in young people is fondamental. By inspiring students and encouraging a scientific way of thinking, astrophysics may contribute to the development of a more informed and critically engaged citizenry, which is essential for addressing complex global challenges. \\

{\it  "For students, physics isn't necessarily what attracts them most these days. But astrophysics is what continues to attract them. And often physics students, even if they don't all end up becoming astrophysicists, are attracted by this research and it helps develop a scientific spirit that is sometimes lacking at the moment. 
So there you go, I can also see that, once again, we can study very specific areas where it could be interesting for studies on climate, on planetary atmospheres or things like that,
I'm not sure that it's where it has the most concrete impact, but I think it's important to maintain, let's say, scientific curiosity and ways of thinking."}

\subsection{Astronomy to question our place in the Universe}
The role of astronomy in questioning man's place in the universe is mentioned. The frequency and insistence with which its importance is underlined is remarkable. Its perceived significance is such that the consequences it may have on education and environmental crisis awareness are sometimes mentioned. 
\\

{\it "One thing I'm becoming increasingly aware of... when I give talks in schools or for associations, for example, when I'm asked to give a talk to the general public, is the... the strength of people's need to, let's say, to... to know where they stand. And when I think about the history of astrophysics, we know that prehistoric man was already observing the stars, and we can think of the Carnac alignments and so on. I think there's... I feel that among children in classes, for example, as well as in the history of the discipline... uh... or in conferences for the general public with adults, there's a strong need to know where we are in the Universe, and I see a link that's quite... not... perhaps strong with this new awareness of the fact that we're on a small planet that's not at all infinite as we... well, we knew it wasn't infinite, but we thought its resources were infinite and that we weren't going to come up against those limits any time soon." }

\section{Consequences for astronomy of substantial reductions of its green house gas emission}
When asked about the climate crisis, all interviewees express their deep concern and the need to react appropriately. Whereas there is 
a diversity of  knowledge regarding the crisis, its consequences and the actions that must be undertaken, 
are  generally advanced and accurate.

\subsection{Reduction of GHG emissions at the individual level}
 At the personal level, both at work or in their private life, research-participants described  a variety of attitudes, including adjustment in housing (improved insulation, use of heat pumps, etc.), dietary changes (local or vegetarian consumption, reduced meat consumption), adoption of waste management practices, change in modes of transportation (from reduced travel to use of electric bicycles and changes in shopping habits). \\

{\it "Yes, that's right, I'd say on both points, on my professional activities and on my lifestyle it's what I've... what I think I understand about climate change that's led me to make changes. In my professional life, I've started to... much less, so I don't buy many new things, just to reduce my overall consumption, my contribution to overall consumption, so I almost only buy used things and I repair them. I haven't stopped eating meat, but I've reduced it enormously. And um... I hardly ever fly on vacation any more. For my leisure trips, I've had to do it once in 5 years. As for my professional activity, I cycle to work as much as I can to avoid using the car every day. I still own a car because I'm the father of 3 children and without a car I haven't yet found any good alternatives for taking my three young children with me when I travel around France. On the other hand, in terms of my professional activity, I'm still... the reduction isn't drastic, because I'm still going to conferences by plane, even if for short journeys, what I'd call short, which are therefore in France and Europe, I now take the train on journeys where I used to take the plane, even if it's longer, within reason, let's say that I haven't made any train journeys lasting more than 6 or 7 hours."}

\subsection{Perceptions and consequences of GHG reductions for astronomy}
When it comes to the possibility of substantially reducing GHG emissions on a disciplinary scale, reluctance is often expressed. The importance of 
fundamental research is expressed and the point is often made that it should not be sacrificed. It is sometimes argued that the GHG emissions remain 
small, if not negligible, compared to other human activities.  It is sometimes argued that  it should not be the research in astronomy doing the largest efforts to reduce GHG emission but, for example, the company presidents or the richest persons. \\

{\it "Well, that would be a shame! (laughs), it would be a shame... Because we're here to try and make progress, to accumulate knowledge, so uh... cutting back on research is a bit of a shame. And then, well, reducing travel is true for... I'd say for people who have permanent jobs, but it's also important for young people to travel. We've seen that during the two years when they've hardly been able to travel, it's been difficult for them to create links with people, to meet people from other countries, and that's important too."  } \\

{\it “Should we stop science and fundamental science for ecological reasons? That's more the question I'm asking. Should we stop sending satellites to study things that we study, which I find extremely important and interesting. They don't have a direct interest in saving the planet, but science in general, for me, is capable of doing that. So I think we shouldn't... Well, how should I put it... At the moment, I think it's unforgivable to stop all fundamental sciences just... I wouldn't say just to save the planet but... How to put it? We can't stop everything. We need to slow down, we need to consume less...”}

\subsection{Possible mitigation plans and solutions}
Several  research-participants believe that significant reductions are necessary and a better use of the existing is often mentioned. 
This typically consists in exploiting more deeply data both from observations and simulations, as well as existing instruments. It is also suggested that 
new algorithms and new approaches will have to be developed.  \\

{\it "After that, we can imagine missions, perhaps, that pool more... Right now, we have missions that are very divided by sector, maybe we could group them together more and try to do this more intelligently... And yeah, the creation of new observatories... Yes, I think we'll have to limit that after a while and what we were saying, make better use of the resources already present because there's already so much to do." } \\

{\it "You can see the image behind here, of my simulations. But in fact we've produced an enormous amount of data with these simulations, most of which is stored and accessible, and which we haven't necessarily exploited to the full. So there's already a huge amount to be done, both for simulation work and for observation. We have an enormous amount of data to work with. So there's already a part of our activity that will perhaps be refocused on exploiting existing data. And then, I think there will be a series of best practices to put in place, including some of the very costly simulations we're doing, which may not be necessary or the right size. There will also be questions that we won't be able to answer any more, and where clearly, in order to push certain questions that we have today in certain fields, we'll have to look for stronger calculation methods, or else become mathematicians of genius and find very, very skilful solutions, which for the moment is not emerging. So, unfortunately, we'll probably have to forget about these questions for a while. From my point of view, I don't see any particularly promising way of answering them, unless we go back to... becoming a mathematician-physicist rather than a numericist, but I'm not even sure that this would answer those questions anyway. So I think part of the answer lies in exploiting existing data". }

\section{Conclusions}
%%--------------------
With the goal of understanding how astronomers perceive the trade-off that scientists are presently facing between the usefulness of 
their activities and the need to reduce the carbon footprint of our societies, we have carried out a series of 28 semi-structured interviews of professional 
astronomers. Astronomy is most of the time seen as an activity that is useful for the society mainly, though not exclusively, at the educational level but also 
regarding the fascination it could exert on at least a fraction of the general public. Most research-participants believe in the IPCC conclusion and fear  global warming and its 
future consequences. They have generally adopted solutions such as flight reduction or alternative transportation to car commute. The question of significant 
GHG reduction at the scale of the community appears much more controversial, in particular regarding the possibility to reduce the numbers of large facilities such as telescopes and satellites. The need to better exploit the existing instruments and data but also to develop new approaches is stressed. Altogether we conclude that by highlighting 
resistances and opportunities,  semi-structured interviews appear to be a powerful tool to help the scientific communities in making choices  in response
to the environmental crisis.

% Optional acknowledgements
% -------------------------
\begin{acknowledgements}
We warmly thank Antoine Hardy for his guidance and his help.
\end{acknowledgements}

\bibliographystyle{aa}  % A&A bibliography style file (aa.bst)
\bibliography{Surname_SXX} % your references in file: Yourfile.bib

\end{document}